\documentclass[a4paper,12pt]{article}

\usepackage{arxiv} 

\usepackage[english]{babel}
\usepackage[utf8]{inputenc} 
\usepackage[T1]{fontenc}    
\usepackage{amsmath}
\usepackage{amssymb} 
\usepackage[font=small,labelfont=bf,labelsep=period]{caption}  
\usepackage{subfig} 
\usepackage{tikz}

\usepackage{graphicx} 
\graphicspath{ {./img/} } 

\usepackage{secdot} 
\sectiondot{subsection} 

\usepackage{multicol} 
\usepackage{float} 

\usepackage[normalem]{ulem}
\usepackage{xcolor}

\usepackage{natbib}
\title{Statistical Properties of Turbulent Fluctuations Associated with Electron-only Magnetic Reconnection}

\author{G. Arrò\\ Department of Mathematics, KU Leuven, Leuven, Belgium\\ Dipartimento di Fisica "E. Fermi", Università di Pisa, Pisa, Italy 
\And F. Califano\\ Dipartimento di Fisica "E. Fermi", Università di Pisa, Pisa, Italy 
\And G. Lapenta\\ Department of Mathematics, KU Leuven, Leuven, Belgium}

\begin{document}
\maketitle

\begin{abstract}

Context: Recent satellite measurements in the turbulent magnetosheath of Earth have given evidence of an unusual reconnection mechanism that is driven exclusively by electrons. This newly observed process was called electron-only reconnection, and its inter-play with plasma turbulence is a matter of great debate.
\\Aims: By using 2D-3V hybrid Vlasov-Maxwell simulations of freely decaying plasma turbulence, we study the role of electron-only reconnection in the development of plasma turbulence. In particular, we search for possible differences with respect to the turbulence associated with standard ion-coupled reconnection.
\\Methods: We analyzed the structure functions of the turbulent magnetic field and ion fluid velocity fluctuations to characterize the structure and the intermittency properties of the turbulent energy cascade.
\\Results: We find that the statistical properties of turbulent fluctuations associated with electron-only reconnection are consistent with those of turbulent fluctuations associated with standard ion-coupled reconnection, and no peculiar signature related to electron-only reconnection is found in the turbulence statistics. This result suggests that the turbulent energy cascade in a collisionless magnetized plasma does not depend on the specific mechanism associated with magnetic reconnection. The properties of the dissipation range are discussed as well, and we claim that only electrons contribute to the dissipation of magnetic field energy at sub-ion scales.

\end{abstract}

\section{Introduction} 

The study of turbulence in a collisionless plasma is an extremely challenging problem to face because it is a strongly nonlinear process involving many decades of scales that extend from fluid magnetohydrodynamic (MHD) scales to ion kinetic and electron kinetic scales that are associated with different physical regimes. No general theory is currently capable of describing the full turbulent cascade process in a plasma. On the other hand, different reduced models have been formulated to describe the properties of the turbulent system in a limited range of spatial and temporal scales and in special physical conditions such as in the presence of a strong magnetic field that makes the plasma anisotropic (see, e.g., \citet{00,01,02} and references therein). Thus, the properties of plasma turbulence can be studied in detail only by means of numerical simulations, within the limits of the currently available computational resources \citep{03a,03,04}. Numerical studies are inspired and guided by in situ satellite measurements taken in the solar wind and in the terrestrial magnetosphere. Space plasmas represent natural laboratories for the study of plasma turbulence through extremely accurate spatial and temporal satellite data \citep{06}. It is worth noting that today, space is the only environment where measurements down to electron scales are accessible, as in the case of the Magnetospheric Multiscale (MMS) space mission \citep{07}, and even ion-scale measurements are far more accurate than in the laboratory.

Solar wind studies focusing on the formation of the turbulent cascade at MHD scales have unambiguously demonstrated the fundamental role of low-frequency Alfvén waves in nonlinearly building up the turbulent spectrum \citep{02}. On the other hand, the properties of the turbulent cascade at kinetic scale are not yet fully understood. Energy transfers at sub-ion scales are thought to be driven by nonlinear interaction between relatively high-frequency modes such as kinetic Alfvén waves and whistler waves \citep{08}. Recent studies instead suggest that the development of turbulence at small scales is closely related to magnetic reconnection phenomena developing inside the current sheets that are spontaneously generated by the turbulent MHD dynamics, which create small-scale coherent structures where energy is thought to be dissipated \citep{04,09,09a}. Understanding the nature of kinetic-scale turbulence in plasmas is therefore an open problem.

Observations of reconnection driven by turbulence have been reported in space plasmas \citep{10a,10b,10c}, and recently, satellite measurements of the MMS mission in the turbulent magnetosheath of Earth have given evidence of unusual reconnection events driven only by electrons, while ions were found to be decoupled from the magnetic field \citep{10}. In particular, satellite data show electron-scale current sheets in which divergent bidirectional electron jets were not accompanied by any ion outflow. This situation is quite different from the standard reconnection picture, in which an electron-scale diffusion region is embedded within a wider ion-scale current sheet. For these reasons, these new phenomena were dubbed "electron-only reconnection events" (e-rec from now on). This discovery stimulated great interest first of all because it is not trivial to determine how electron-scale current sheets undergoing e-rec may form in a large-scale turbulent environment. For instance, in the terrestrial magnetosheath, energy is typically first transferred in a continuous way from large MHD scales down to ion kinetic scales (or directly injected by reconnection at ion kinetic scales) and finally to the electron kinetic scale. A fundamental question to answer is therefore how e-rec can be triggered by the turbulent motion of a plasma. This problem has recently been addressed by Califano et al. \citep{11}, who showed using 2D-3V dimensional Eulerian hybrid Vlasov-Maxwell simulations \citep{12a,12} that if the scale of injection of energy in a turbulent plasma is close to the ion kinetic scale,  ions decouple from the magnetic field and reconnection processes taking place in the system are driven exclusively by electrons, showing the same features of the e-rec events detected by MMS. The transition from standard ion-coupled reconnection to e-rec has recently been studied in detail by Pyakurel et al. \citep{12b} using 2D-3V dimensional particle-in-cell simulations of laminar reconnection with conditions appropriate for the magnetosheath. By gradually increasing the size of the simulation box from a few ion inertial lengths to several tens of ion inertial lengths, they observed a smooth transition from the e-rec regime, where ions are decoupled from the reconnection dynamics, to the more familiar ion-coupled reconnection. 

Another important aspect concerning the relationship between e-rec and turbulence is to understand whether and how this new reconnection process in turn affects the development of the turbulent energy cascade and its statistical properties, and if there are any differences with respect to the turbulence associated with standard reconnection. In this context, the magnetosheath data collected by the MMS satellites were recently analyzed by Stawarz et al. \citep{12c}, who showed that the statistical distribution of the turbulent magnetic field fluctuations associated with e-rec and their spectral properties are analogous to those observed in other turbulent plasmas, such as the solar wind, and in numerical simulations of plasma turbulence. 

In this paper we present a study of the statistical properties of fluctuations developing in a simulation of freely decaying plasma turbulence in which e-rec occurs. The results obtained from this simulation are then compared to those of a different simulation of plasma turbulence where standard reconnection takes place. We aim at finding possible differences between the statistical features of these two turbulent systems by taking advantage of the different dynamics of the ions  associated with the reconnection structures in the two simulations. In particular, we investigate if there is any specific signature of e-rec in the turbulence statistics. Our study is based on the analysis of the structure functions (hereafter, SFs) of turbulent fields. SFs have been used extensively to analyze numerical simulations \citep{12d,12d2} and observational data \citep{12e}, showing that the turbulent magnetic field undergoes a transition from an intermittent dynamics to a self-similar one at sub-ion scales \citep{12d,12e}. Here we extend the SFs analysis to ion velocity fluctuations as well in order to characterize the behavior of this species, which has a very different role in the reconnection dynamics of the two simulations. Our main finding is that the turbulent fluctuations associated with e-rec show the same statistical properties as the turbulent fluctuations associated with standard ion-coupled reconnection. The structure of the turbulent cascade is also examined. In particular, the properties of the magnetic field dissipation range of a collisionless turbulent plasma are discussed, and we claim that only electrons contribute to its formation.

The paper is structured as follows: the numerical model implemented in our simulations is discussed in section 2. In section 3 we describe the specific setup adopted for the two simulations considered here, which are the same as were analyzed by Califano et al. (2020). The method of analysis based on the study of SFs is introduced in section 4, and our results are presented in section 5. Our conclusions are finally discussed in the last section.

\section{Numerical Model} 

The two simulations analyzed in this paper were performed using an Eulerian hybrid Vlasov-Maxwell (HVM) 2D-3V dimensional code that advances the Vlasov equation for ions in time \citep{12a}, coupled with an isothermal fluid model with finite mass for the electrons \citep{12}. The electron response is described by the generalized Ohm law that includes electron inertia terms, allowing the complete decoupling of the magnetic field at electron scales \citep{12},\\
\begin{gather}
\textbf{E}-\frac{d^2_e}{n}\nabla^2\textbf{E}=\frac{1}{n}(\textbf{J}\times\textbf{B})-(\textbf{u}\times\textbf{B})-\frac{1}{n}\nabla P_e + \notag \\
\\
+\frac{d^2_e}{n} \nabla \cdot \left(\textbf{u}\textbf{J}+\textbf{J}\textbf{u}- \frac{\textbf{J}\textbf{J}}{n} \right), \notag
\end{gather}
\\where $\textbf{E}$ and $\textbf{B}$ are the electric and magnetic fields, respectively, $\textbf{J}\!=\!\nabla \times \textbf{B}$ is the current density (we neglect the displacement current), $\textbf{u}$ and $P_i$ are the ion velocity and pressure, respectively, $P_e\!=\!nT_e$ is the electron isothermal pressure, and $d_e$ is the electron inertial length. Furthermore, quasi-neutrality is assumed so that ion and electron densities are equal $n_e\!=\!n_i\!=\!n$. Finally, the evolution of the magnetic field is described by the Faraday equation. All equations were normalized and transformed in dimensionless units using the ion mass $m_i$, charge $+e$, inertial length $d_i$ , and cyclotron frequency $\Omega_i$ (see \citet{12}). For the sake of numerical stability, a numerical filter that smooths out the electromagnetic fields at high wave numbers was used \citep{13}.

\section{Simulations Setup}

We report two simulations that were identical in every aspect except for the spectrum of modes that was used to initially drive turbulence: the first simulation was initialized with ion-scale fluctuations so that e-rec can take place, with ions that do not participate significantly, whereas the second simulation has only large-scale perturbations and reconnection occurs in the usual ion-coupled regime.

The equations of the HVM model were integrated on a 2D-3V domain (bidimensional in real space and tridimensional in velocity space). In both simulations we took a square spatial domain of size $L\!=\!20 \pi d_i$ covered by a uniform grid consisting of $1024^2$ mesh points, while the velocity domain was cubic with sides spanning from $-5v_{th,i}$ to $+5v_{th,i}$ in each direction (where $v_{th,i}$ is the ion thermal velocity) and sampled by a uniform grid consisting of $51^3$ mesh points. The ion-to-electron mass ratio was $m_i/m_e\!=\!144,$ which implies $d_i/d_e\!=\!\sqrt[]{m_i/m_e}\!=\!12$, the electron temperature was set to $T_e\!=\!0.5,$ and the plasma beta was $\beta\!=\!1$, corresponding to $v_{th,i}\!=\!\sqrt[]{\beta/2}\!=\!\sqrt[]{1/2}$ (in Alfvén speed units). Both simulations were initialized with an isotropic Maxwellian distribution for ions and an homogeneous out-of-plane guide field $\textbf{B}_0$ along the $z$ -axis. Turbulence was triggered by adding to the guide field some large-scale, random phase, isotropic magnetic field sinusoidal perturbations $\delta \textbf{B}$. For the first simulation (hereafter, sim.1), we took perturbations with wavenumber $k$ in the range $0.1 \leqslant k d_i \leqslant 0.6$, mean amplitude $\left| \delta\textbf{B} \right|_{rms}/B_0\!\simeq\!0.2,$ and maximum amplitude $\left| \delta\textbf{B} \right|_{max}/B_0\!\simeq\!0.5$. The scales of the largest wavenumbers of these perturbations were close to ion kinetic scales in order for the ions to be nearly decoupled from the magnetic field dynamics from the beginning of the simulation and therefore to drive a turbulent environment in which e-rec occurs \citep{11}. In the second simulation (hereafter, sim.2), the system was perturbed by fewer modes with wavenumber $k$ in the range $0.1 \leqslant k d_i \leqslant 0.3$, all being far larger than ion kinetic scales, mean amplitude $\left| \delta\textbf{B} \right|_{rms}/B_0\!\simeq\!0.25,$ and maximum amplitude $\left| \delta\textbf{B} \right|_{max}/B_0\!\simeq\!0.5$. In this way, ions were magnetized at the beginning of the simulation, eventually leading to a turbulent environment in which standard reconnection occurs. The time step used for both simulations was $\Delta t\!=\!0.005 \,\Omega^{-1}_i$ in order to accurately resolve phenomena with frequencies between the electron cyclotron frequency $\Omega_e$ and the ion cyclotron frequency $\Omega_i$. This choice is consistent with the limits of our HVM model, where ions are kinetic, while electrons, with mass, are taken as fluid but adopting the Ohm law corresponding to an electron magnetohydrodynamics (EMHD) dynamics \citep{11}. It is worth noting that in our simulations, with the spatial resolution we chose, we have only two points to resolve the electron inertial length $d_e$ , but nonetheless this was sufficient to distinguish the EMHD invariant $F\!\doteq\!\psi-d_{\rm e}^2\nabla^2\psi$ from the flux function $\psi$ \citep{23}, in other words, it allowed us to accurately resolve the electron physics at sub-ion scales (see \citet{11} for a detailed discussion of this point).

We did not include any external forcing term in our model (in this case, we talk about freely decaying turbulence simulations). This means that when the plasma is in a turbulent regime, the energy dissipated at small scales  is not replaced by any large-scale energy source and the system will never reach the statistically stationary state corresponding to a fully developed turbulence \citep{14}. However, there is a time interval during which 2D freely decaying turbulence reaches a peak of activity that shows statistical properties that are very similar to those of homogeneous and isotropic fully developed turbulence. This time interval corresponds to a period in which the out-of-plane mean square current $\langle J^2_z \rangle$ reaches and maintains a roughly constant peak value \citep{03a,04,12d} that corresponds to an intense small-scale activity \citep{16}. For this reason, the analysis of the turbulence statistics was carried out at a fixed time close to the peak of $\langle J^2_z \rangle$ in both simulations.

\section{Structure functions and intermittency} \label{sec04}

One way to characterize a turbulent process, regardless of its nature, is to analyze the statistical features of the fluctuations of the physical quantities at different scales \citep{01,14}. For a plasma, these quantities could be for example the magnetic field $\textbf{B}$ or the ion velocity $\textbf{u}$. Given a generic vector quantity $\textbf{q}(\textbf{x})$, its fluctuations in the direction of $\textbf{r}$ at scale $r\!=\!|\textbf{r}|$ can be defined as \citep{14}\\
\begin{equation}
\Delta q_\parallel(\textbf{x},\textbf{r})\!=\![\textbf{q}(\textbf{x}+\textbf{r})\!-\!\textbf{q}(\textbf{x})]\!\cdot\!\frac{\textbf{r}}{r}
.\end{equation}
\\The moments of order $p$ of such fluctuations are given by\\
\begin{equation}
S_p(\textbf{r})\!=\!\langle |\Delta q_\parallel(\textbf{x},\textbf{r})|^p \rangle \label{sf}
\end{equation}
\\and are known as (longitudinal) structure functions of the variable $\textbf{q}(\textbf{x})$, where the symbol $\langle \cdot \rangle$ indicates the average on a suitable statistical ensemble. For a homogeneous and isotropic system, SFs depend solely on $r$ and the ensemble average can be replaced by an average over the real space. 

The importance of SFs in analyzing turbulence lies in the fact that in many turbulent processes, they take the form of a power law,\\
\begin{equation}
S_p(r)\sim r^{\xi(p)} \label{power}
,\end{equation}  
\\where $\xi(p)$ is called the scaling exponent of the process. This exponent contains important information about the spatial distribution of fluctuations. It is possible to prove that if $\xi(p)\!=\!ph$ (with $h$ being a constant), the fluctuations are self-similar, that is, they are uniformly distributed in the system at all scales. On the other hand, if $\xi(p)$ is nonlinear in $p,$ the fluctuations are intermittent, which means that they become increasingly less homogeneous with decreasing scale length, and they tend to be concentrated only in some portions of the system \citep{14}. Therefore SFs represent a powerful analysis tool that allows us to identify some key properties of a turbulent process.

Sometimes the SFs of finite systems where turbulence is not fully developed do not take the form of the power law of eq. (\ref{power}). Nevertheless, the turbulent flow can still be characterized using a set of scaling exponents $\xi(p)$ if by plotting SFs of different order one against the other, the following scaling is obtained:\\
\begin{equation}
S_p(r) \sim S_q(r)^{\beta(p,q)}
,\end{equation}
\\where $\beta(p,q)\!=\!\xi(p)/\xi(q)$. In this case, we talk about extended self-similarity (ESS), which has been observed in many experimental turbulent systems as well as in numerical simulations \citep{17,18,19}. In the case of ESS, it is not possible to calculate all the $\xi(p)$ separately because they appear in the form of a fraction in the scaling exponents $\beta(p,q)$. However, the knowledge of $\beta(p,q)$ alone is sufficient to determine whether the turbulent cascade is self-similar or intermittent. In the case of self-similarity, $\xi(p)\!=\!ph$ and so $\beta(p,q)\!=\!p/q$, while in the case of intermittency, $\beta(p,q)\!\neq\!p/q$ \citep{12d}.

In our case, the SFs were calculated by assuming homogeneity and isotropy in both simulations. In this way, eq. (\ref{sf}) reduces to\\
\begin{gather}
S_p(r)=\langle |q_x(x+r,y)-q_x(x,y)|^p \rangle=\notag \\
\\
\langle |q_y(x,y+r)-q_y(x,y)|^p \rangle, \notag
\end{gather}
\\where the ensemble average is replaced by the average over real space. The assumption of homogeneity and isotropy was confirmed by comparing the SFs calculated using $q_x$ with the SFs calculated using $q_y$ , and we found only very little difference between them for all quantities we considered in the two simulations. SFs higher than $p\!=\!4$ were not considered here because calculating them requires a larger simulation grid with many more points in the real space domain than we used \citep{20a,20}. This problem is related to the fact that the calculation of high-order moments of a quantity strongly depends on the tails of its distribution, which are often associated with low probability. As a result, when the ensemble average is replaced with the real space average, it is necessary to ensure that the number of sampled grid points is large enough to include the tail events. 

\section{Results}

The statistical analysis of the turbulent fluctuations in the two simulations was carried out at a fixed time when the turbulent activity was at its maximum. For sim.1, where e-rec is observed, this time corresponds to $t_1\! = \!131.7 \,\Omega^{-1}_i$ , while for sim.2, where magnetic reconnection develops according to the standard picture, this time corresponds to $t_2\! = \!147.5 \,\Omega^{-1}_i$.

\begin{figure}[t]
\centering
\subfloat{
\begin{tikzpicture}
\node[] at (0,0) { \includegraphics[width=.48\linewidth]{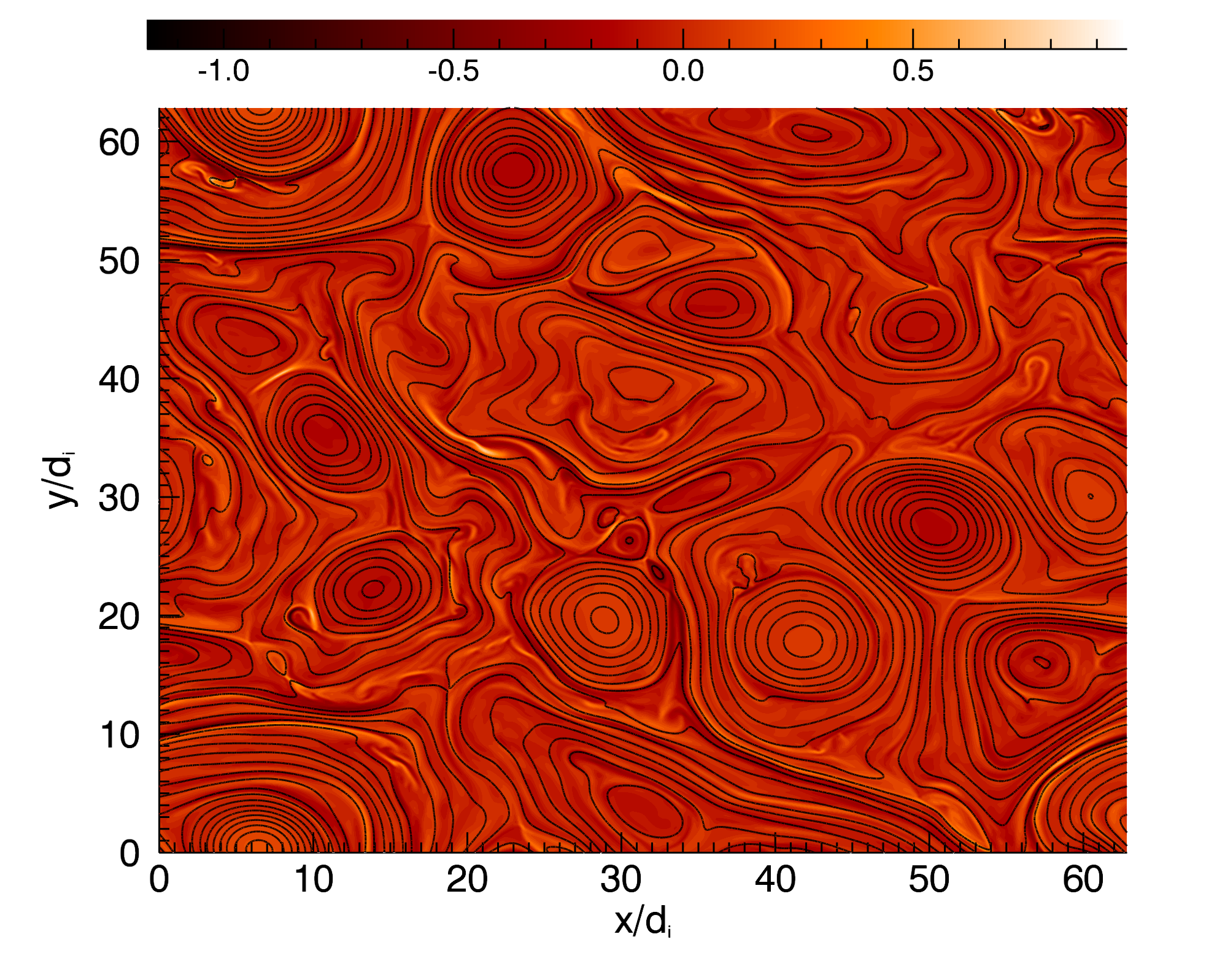} };
\node[] at (-4.2,3) { \Large{(a)} };
\end{tikzpicture}
}
\subfloat{
\begin{tikzpicture}
\node[] at (0,0) { \includegraphics[width=.48\linewidth]{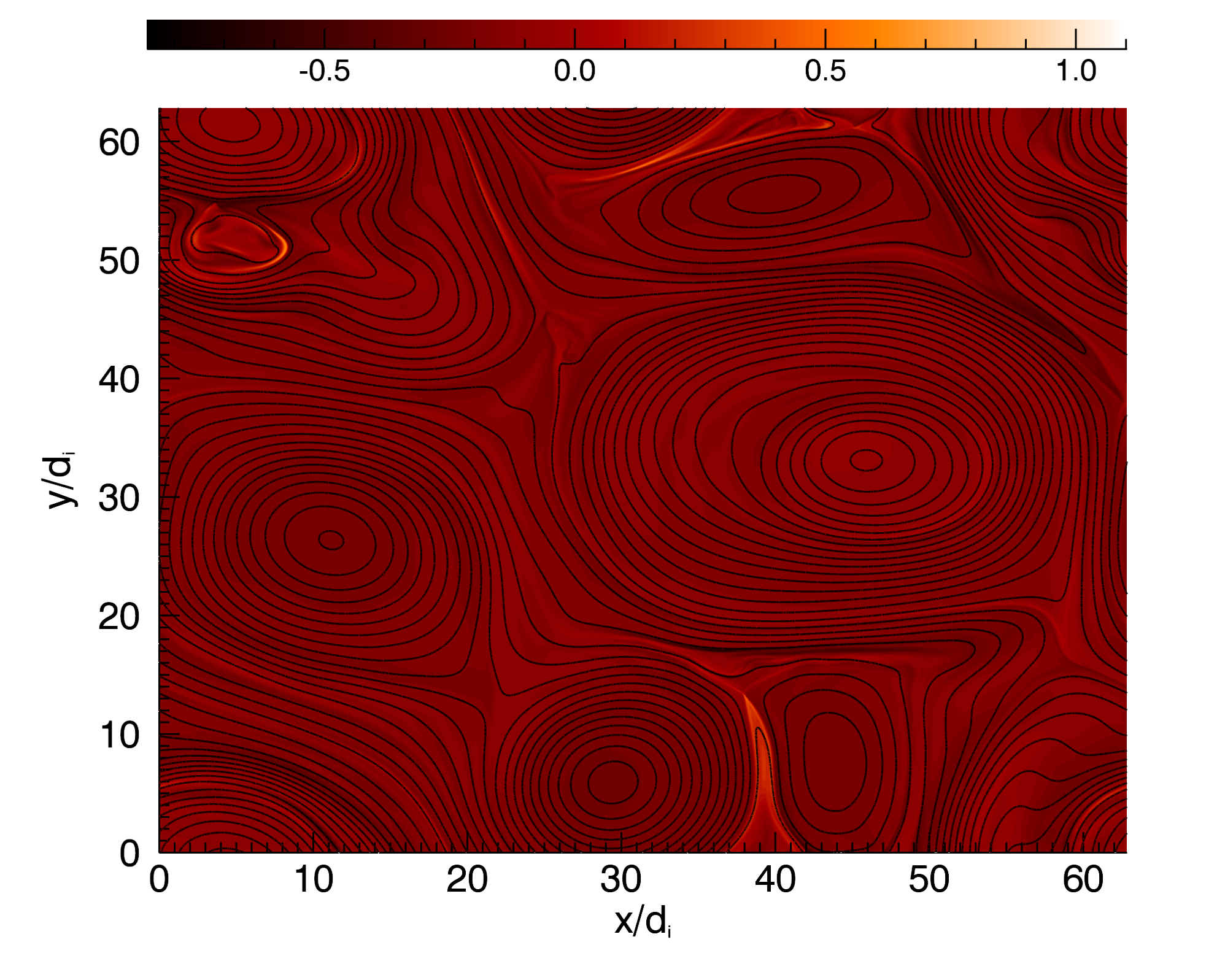} };
\node[] at (-4.2,3) { \Large{(b)} };
\end{tikzpicture}
}
\\
\subfloat{
\begin{tikzpicture}
\node[] at (0,0) { \includegraphics[width=.48\linewidth]{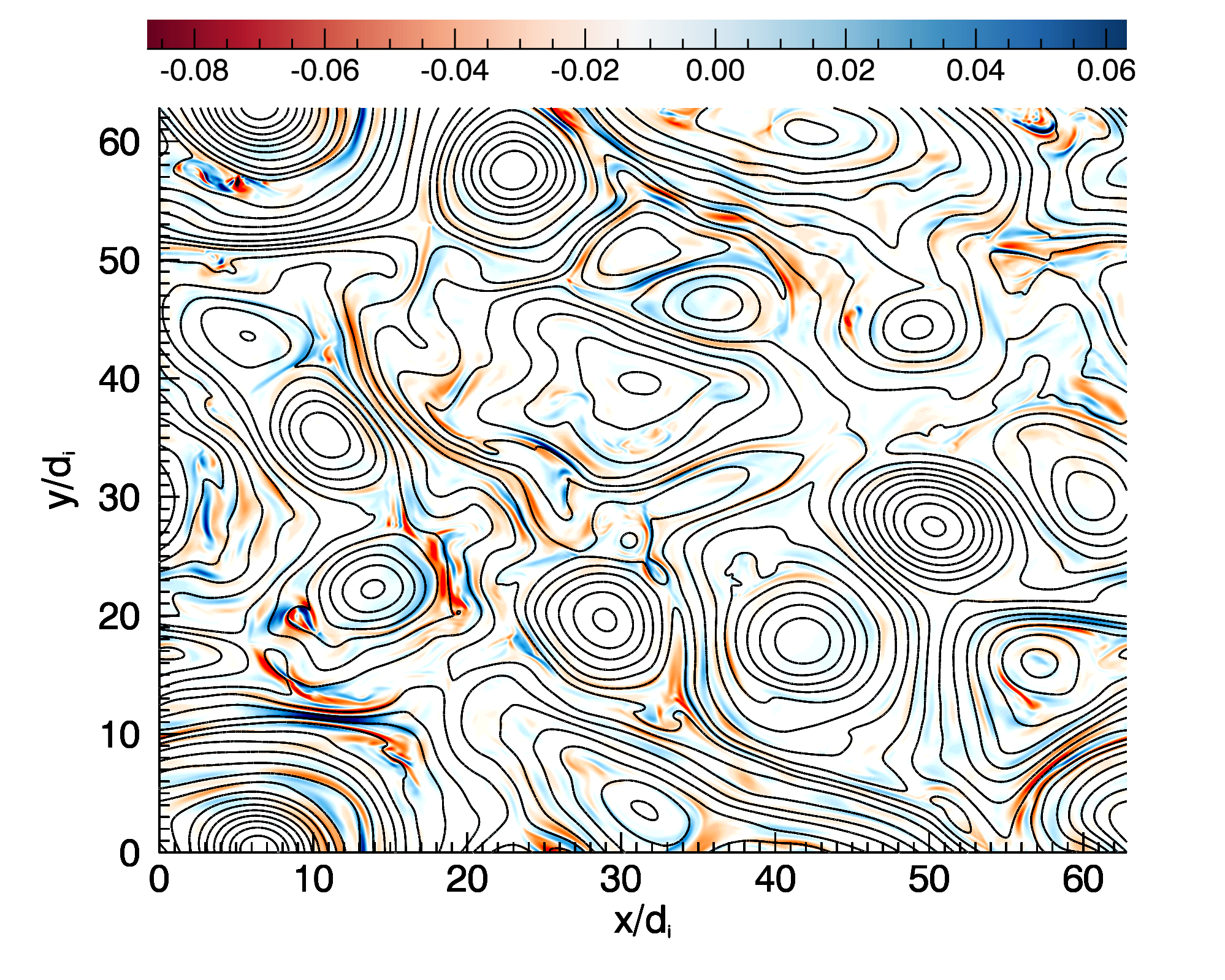} };
\node[] at (-4.2,3) { \Large{(c)} };
\end{tikzpicture}
}
\subfloat{
\begin{tikzpicture}
\node[] at (0,0) { \includegraphics[width=.48\linewidth]{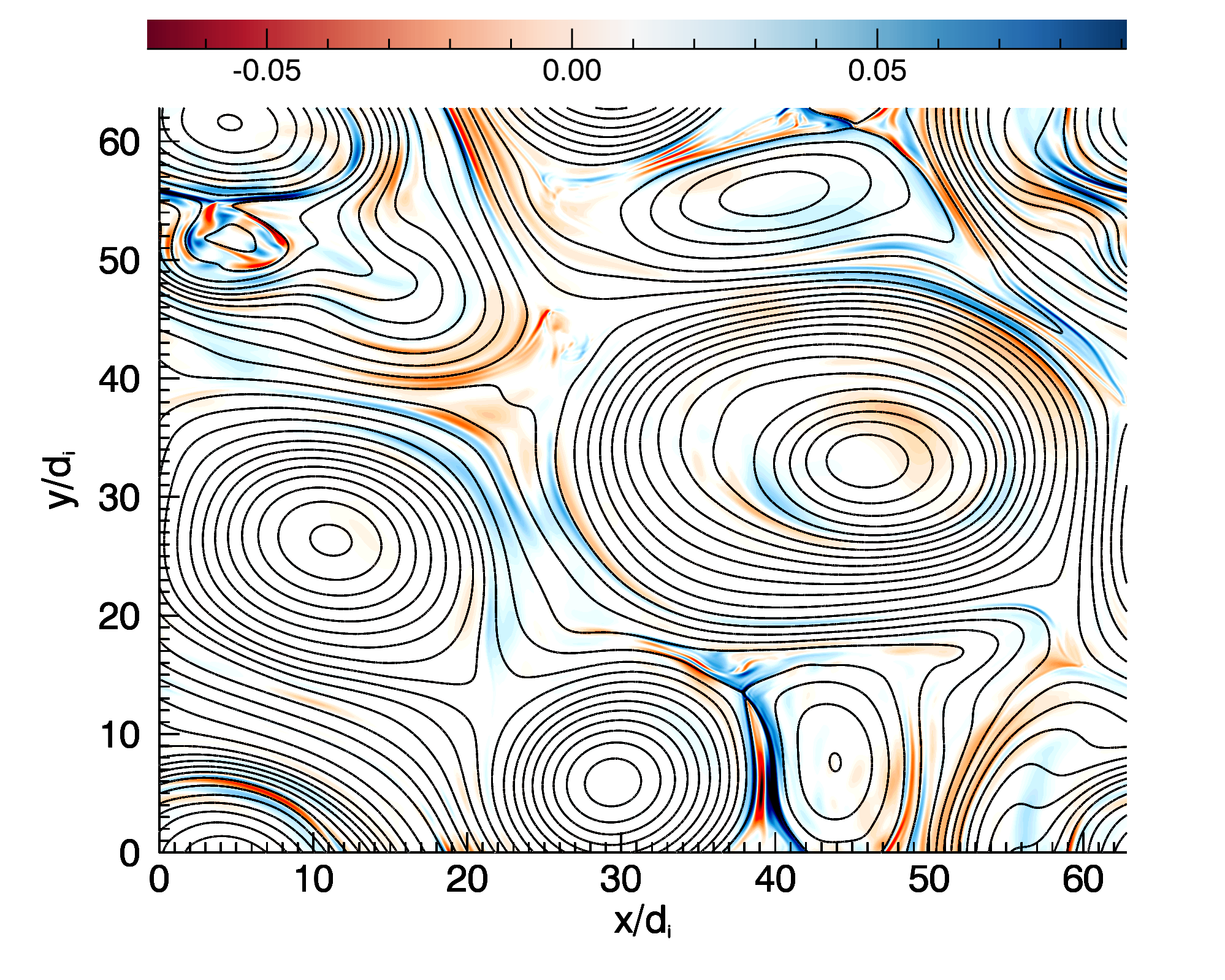} };
\node[] at (-4.2,3) { \Large{(d)} };
\end{tikzpicture}
}
\caption{Top panels: Shaded contour plots of the out-of-plane current $J_z$ (colored) and contour lines of the flux function $\Psi$ (black lines) of sim.1 at $t_1\!=\!131.7 \,\Omega^{-1}_i$ (a) and of sim.2 at $t_2\!=\!147.5 \,\Omega^{-1}_i$ (b). Bottom panels: Shaded contour plots of $\textbf{J}\cdot\textbf{E}$ (colored) and contour lines of $\Psi$ (black lines) of sim.1 at $t_1\!=\!131.7 \,\Omega^{-1}_i$ (c) and of sim.2 at $t_2\!=\!147.5 \,\Omega^{-1}_i$ (d).}
\label{fig_Jz_e-rec}
\end{figure}
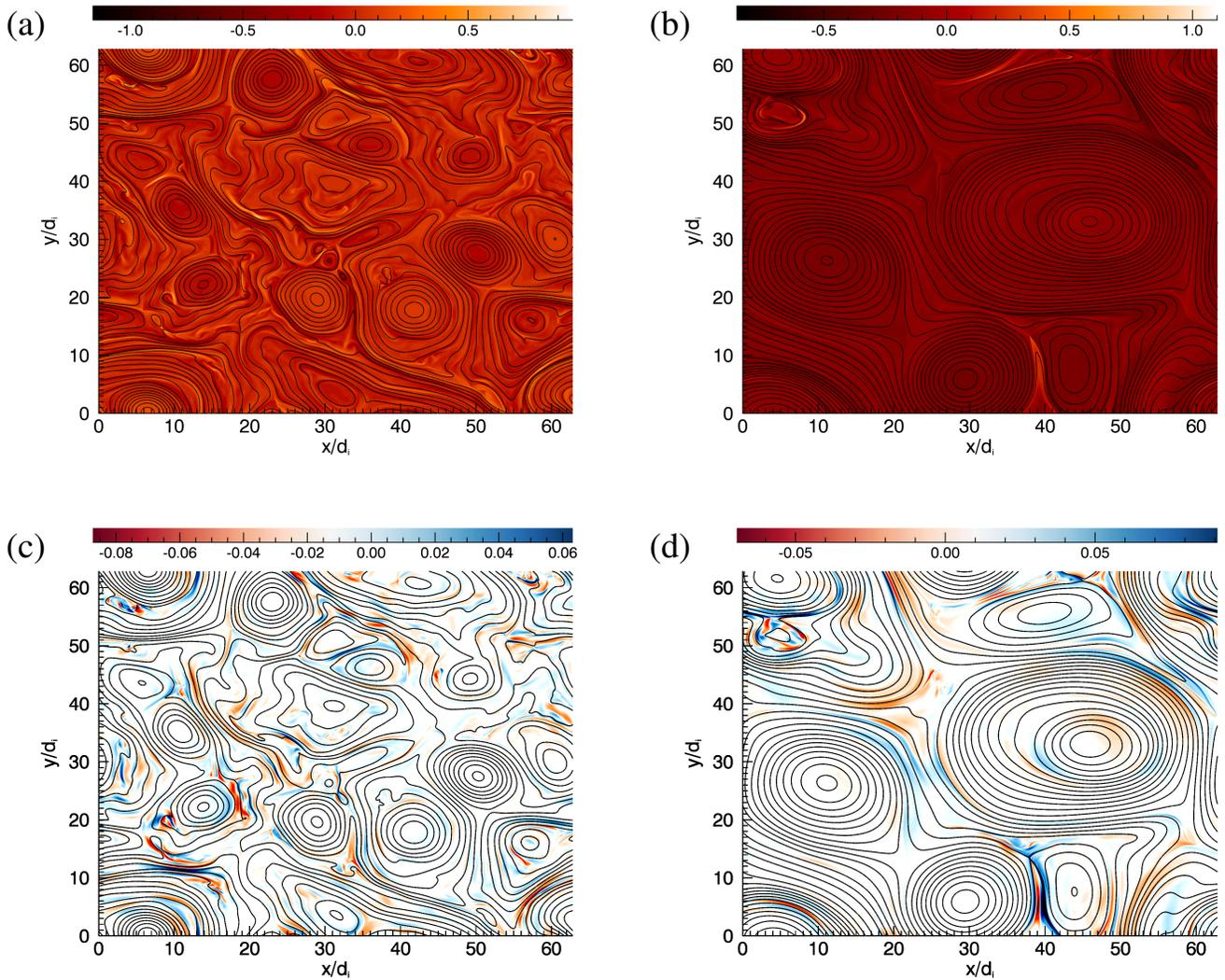

In the top panels of figure \ref{fig_Jz_e-rec} we show for both simulations the shaded contour plots of the out-of-plane current $J_z$ together with the contour lines of the flux function $\Psi$, related to the in-plane magnetic field by $\textbf{B}_\perp\!=\!\nabla\Psi\times\textbf{e}_z$ (with $\textbf{e}_z$ being the out-of-plane unit vector). In both simulations we see that the magnetic configuration of the system is characterized by a large number of island-like magnetic structures of various sizes and shapes, produced by the nonlinear evolution of the initial perturbation. The process of the formation of reconnection sites and the development of an intermittent turbulent cascade of magnetic energy can be understood in terms of the nonlinear interaction between these magnetic islands, which attract one another when the associated central current $J_z$ is of the same sign (and vice versa in the case of opposite sign). In particular, as two islands with central $J_z$ of the same sign approach each other, the magnetic field lines of opposite sign between them are pushed against each other, and this leads to the formation of a thin current sheet where reconnection occurs and magnetic energy is dissipated. Thus, as a result of this dynamics, reconnecting current sheets are not uniformly distributed in a turbulent plasma, they tend to be concentrated between merging magnetic islands, and therefore the dissipation of magnetic energy is nonuniform, that is, the turbulent cascade of magnetic energy is intermittent. The relation between the formation of localized reconnecting current sheets and the development of an intermittent turbulent cascade is highlighted by the contour plots of $\textbf{J}\cdot\textbf{E}$ shown in figure \ref{fig_Jz_e-rec} (the flux function $\Psi$ is overplotted), bottom panels, made at the same time instants and for the same runs as the corresponding contour plots of $J_z$ in the top panels. The quantity $\textbf{J}\cdot\textbf{E}$, representing the energy exchange between the electromagnetic field and the plasma, is significantly nonzero only in correspondence to the intense current structures, thus marking the strong correlation between reconnection and the intermittent dissipation of magnetic energy.

The characteristic size of the magnetic islands depends on the wavelength of the initial fluctuations, that is, on the injection scale, and therefore magnetic islands in sim.1 are smaller than those in sim.2. As a result, the characteristic thickness and length of the current sheets in the two simulations are different as well, and this affects the ion magnetization and consequently the dynamics of magnetic reconnection \citep{12b}. It has been shown in Califano et al. (2020) that in sim.1, ions are (and remain) decoupled from the magnetic field on the scale of the current sheets and because of this, e-rec develops. Conversely, the current sheets of sim.2 are large enough to let the ions participate in the magnetic field dynamics, hence reconnection proceeds according to the standard ion-coupled reconnection model. A statistical analysis of the characteristic widths and lengths of the current structures of the two simulations here considered has been carried out by Califano et al. (2020),   who showed that the reconnecting current sheets of sim.1 are shorter than those of sim.2, while their characteristic width is about the same in the two simulations. In particular, in sim.2 the characteristic length of the current sheets was found to be at least about $10 d_i$ and to vary up to scales of some tens of $d_i$. On the other hand, in sim.1, all the reconnecting current sheets have about the same length, which is about a few $d_i$.

In summary, the turbulent magnetic fluctuations of sim.1 and sim.2 have a significantly different local dynamics. We now determine whether there is a difference in their statistical features, in particular by analyzing the SFs of the magnetic field.

\begin{figure}[t]
\centering
\subfloat{
\begin{tikzpicture}
\node[] at (0,0) { \includegraphics[width=.48\linewidth]{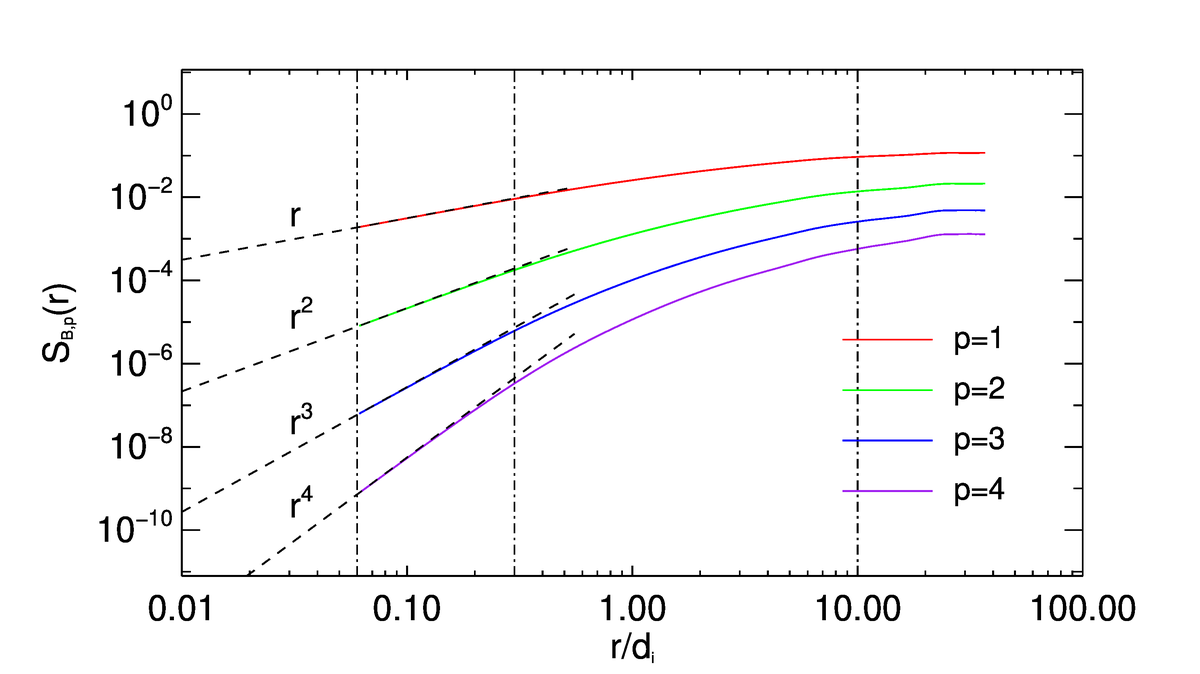} };
\node[] at (-4.1,1.7) { \Large{(a)} };
\node[] at (-1.2,-1.4) { \scriptsize{Range I} };
\node[] at (0.6,-1.4) { \scriptsize{Range II} };
\end{tikzpicture}
}
\subfloat{
\begin{tikzpicture}
\node[] at (0,0) { \includegraphics[width=.48\linewidth]{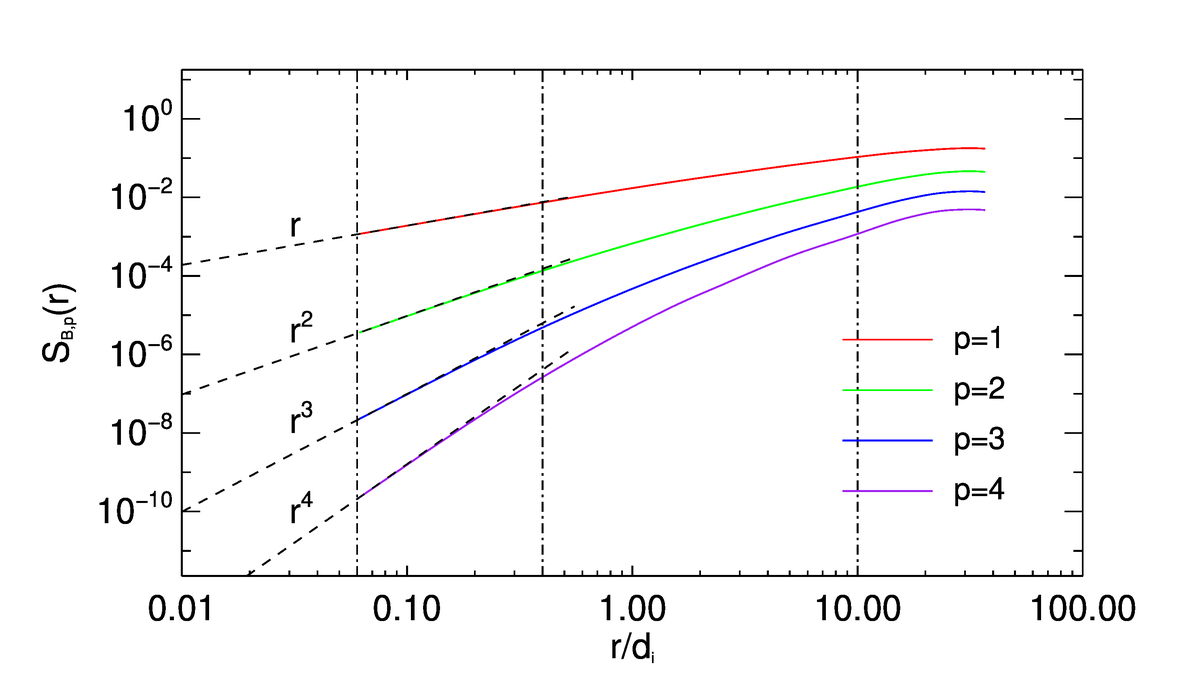} };
\node[] at (-4.1,1.7) { \Large{(b)} };
\node[] at (-1.2,-1.4) { \scriptsize{Range I} };
\node[] at (0.6,-1.4) { \scriptsize{Range II} };
\end{tikzpicture}
}
\\
\subfloat{
\begin{tikzpicture}
\node[] at (0,0) { \includegraphics[width=.48\linewidth]{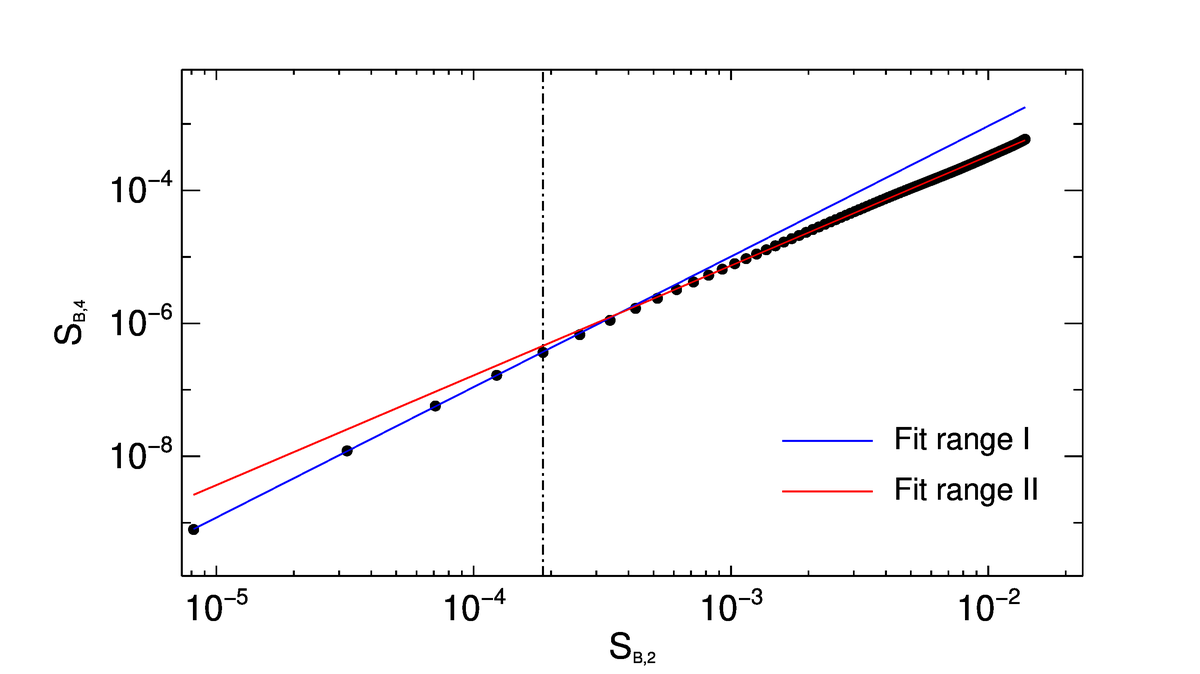} };
\node[] at (-4.1,1.7) { \Large{(c)} };
\end{tikzpicture}
}
\subfloat{
\begin{tikzpicture}
\node[] at (0,0) { \includegraphics[width=.48\linewidth]{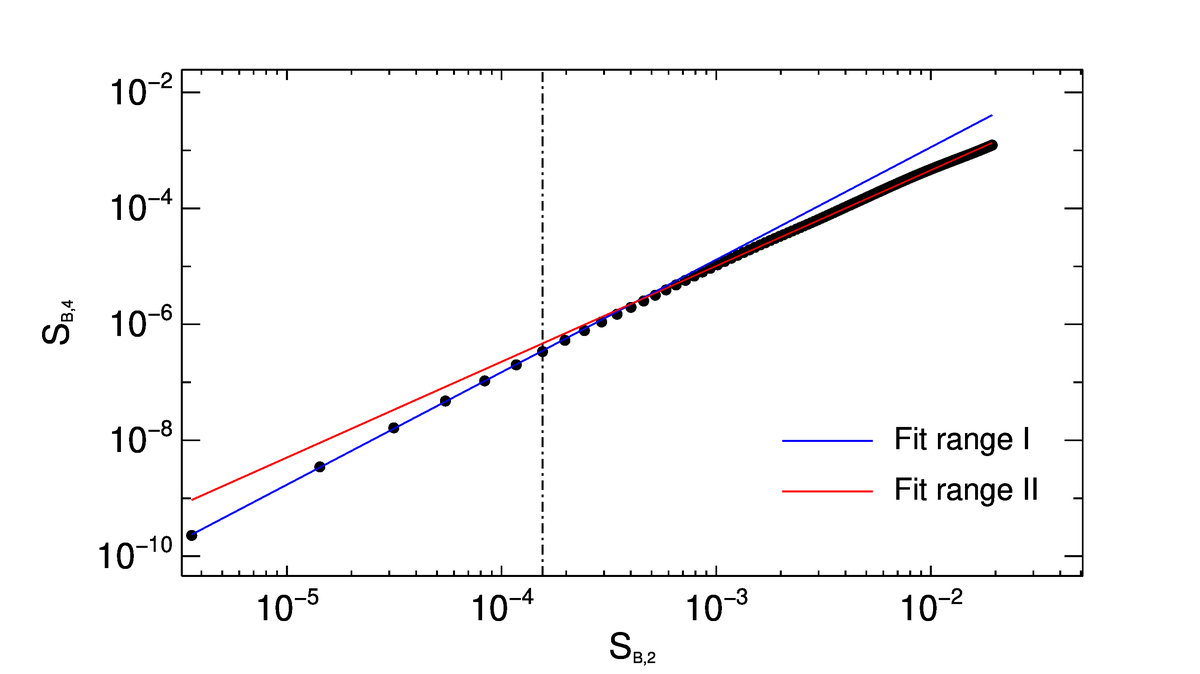} };
\node[] at (-4.1,1.7) { \Large{(d)} };
\end{tikzpicture}
}
\\
\subfloat{
\begin{tikzpicture}
\node[] at (0,0) { \includegraphics[width=.48\linewidth]{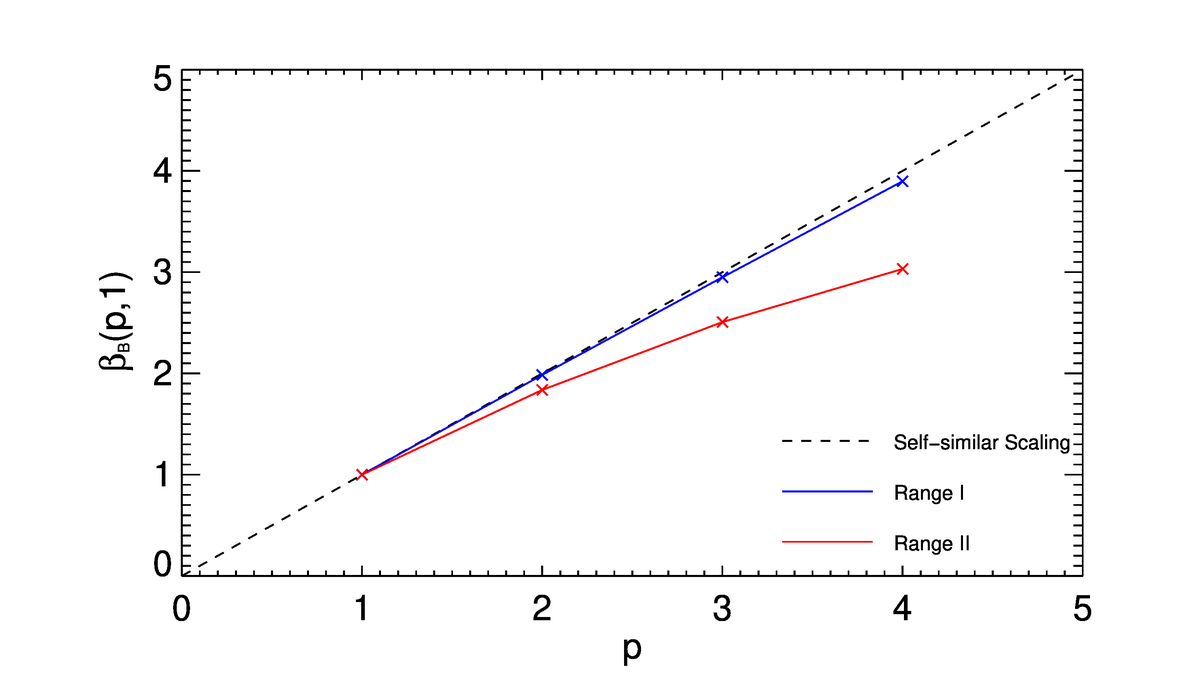} };
\node[] at (-4.1,1.7) { \Large{(e)} };
\end{tikzpicture}
}
\subfloat{
\begin{tikzpicture}
\node[] at (0,0) { \includegraphics[width=.48\linewidth]{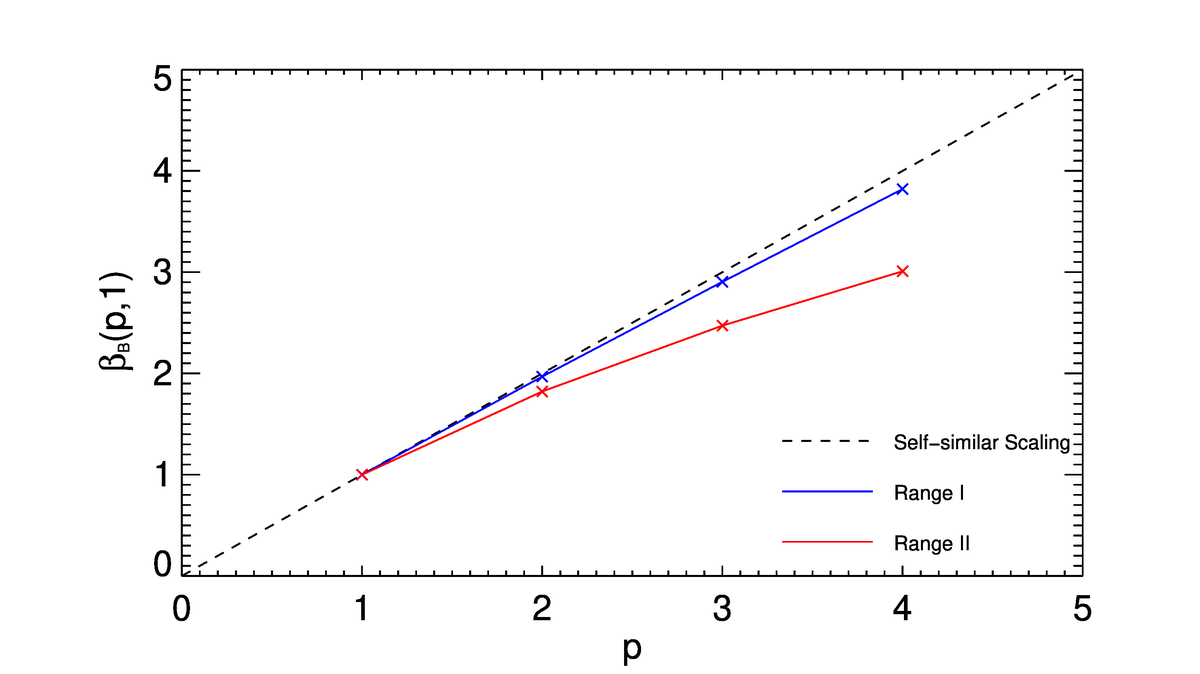} };
\node[] at (-4.1,1.7) { \Large{(f)} };
\end{tikzpicture}
}
\caption{Top panels: Magnetic field structure functions $S_{B,p}$ (log-scale) of sim.1 at $t_1\!=\!131.7 \,\Omega^{-1}_i$ (a) and of sim.2 at $t_2\!=\!147.5 \,\Omega^{-1}_i$ (b); dashed straight lines represent the power laws $r^p$ , and vertical dash-dotted lines delimit ranges I and II. Middle panels: $S_{B,4}$ vs. $S_{B,2}$ (filled black dots, in log-scale) for $r\!<\!10\,d_i$ from sim.1 at $t_1\!=\!131.7 \,\Omega^{-1}_i$ (c) and sim.2 at $t_2\!=\!147.5 \,\Omega^{-1}_i$ (d); ranges I and II were fit separately with straight lines (blue and red lines, respectively); vertical dash-dotted lines separate range I from range II. Bottom panels: Magnetic field scaling exponents $\beta_B(p,1)$ within range I (blue) and range II (red) from sim.1 at $t_1\!=\!131.7 \,\Omega^{-1}_i$ (e) and sim.2 at $t_2\!=\!147.5 \,\Omega^{-1}_i$ (f); the dashed straight line, representing the self-similar scaling $\beta(p,1)\!=\!p$, is given as reference.}
\label{fig02}
\end{figure}
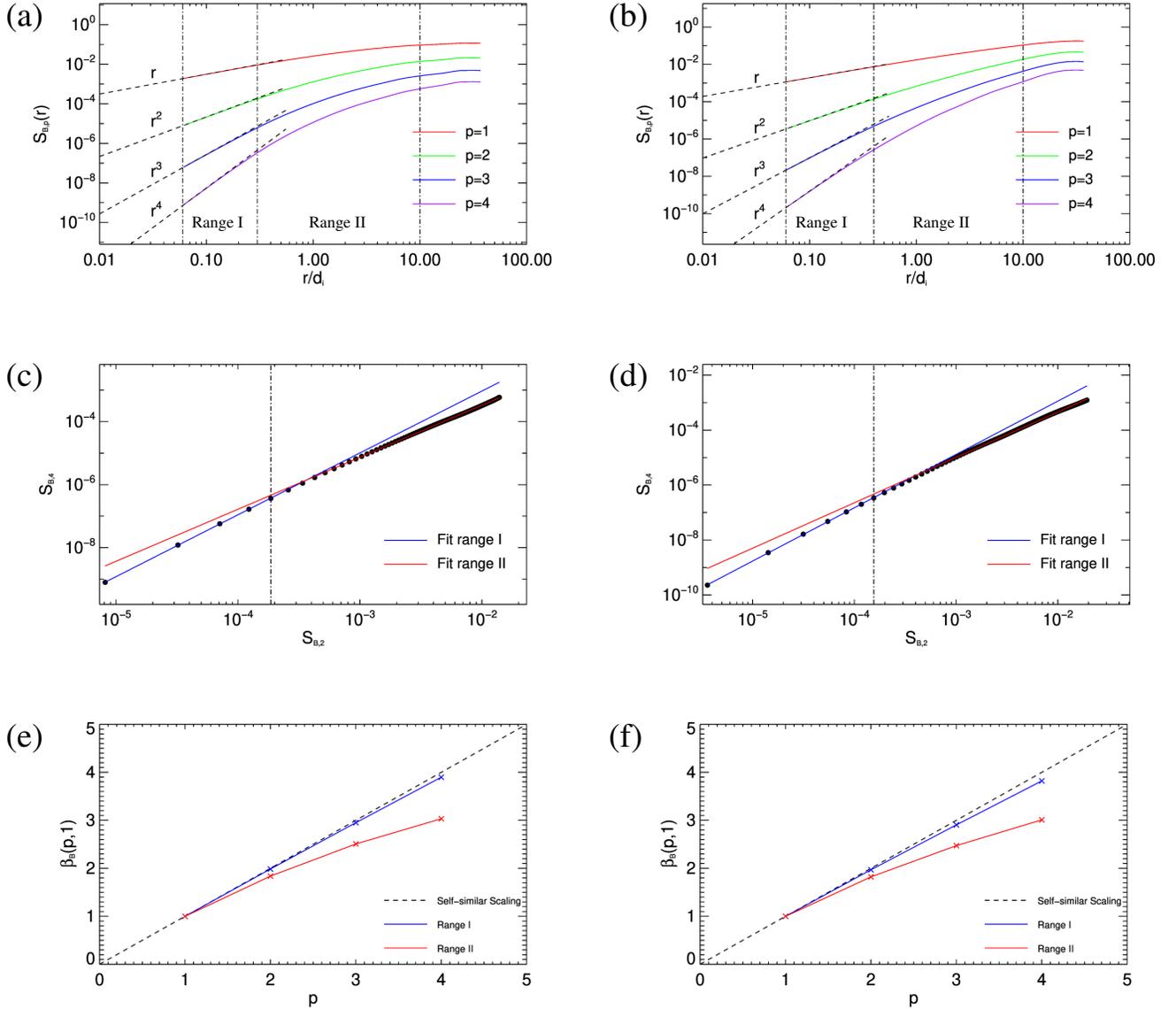

In panels (a) and (b) of figure \ref{fig02} we compare the first four magnetic field structure functions $S_{B,p}$ (in logarithmic scale) of sim.1 and sim.2, respectively. These SFs were calculated using $B_x$. The same results were obtained using $B_y$ (not shown here), which means that magnetic field turbulence is isotropic in our simulations. Figure \ref{fig02} shows that all magnetic field SFs of both simulations have the same behavior over the range of scales we considered and that there are no significant differences between sim.1 and sim.2. In particular, we see that for $r\!>\!10\,d_i$ , all SFs start to saturate, while for $r\!<\!10\, d_i$ , it is possible to distinguish two ranges that correspond to two different scalings. The first range, hereafter called range I, extends from $r\!\simeq\!0.06\, d_i$ to $r\!\simeq\!0.3\, d_i$. Here the SFs follow the power law of eq. (\ref{power}). The second range, hereafter called range II, reaches from $r\!\simeq\!0.3\, d_i$ to $r\!\simeq\!10\, d_i$. In this range, $log(S_{B,p})$ is nonlinear in $log(r),$ which means that the SFs do not take the form of a power law.

\begin{figure}[t]
\centering
\subfloat{
\begin{tikzpicture}
\node[] at (0,0) { \includegraphics[width=.48\linewidth]{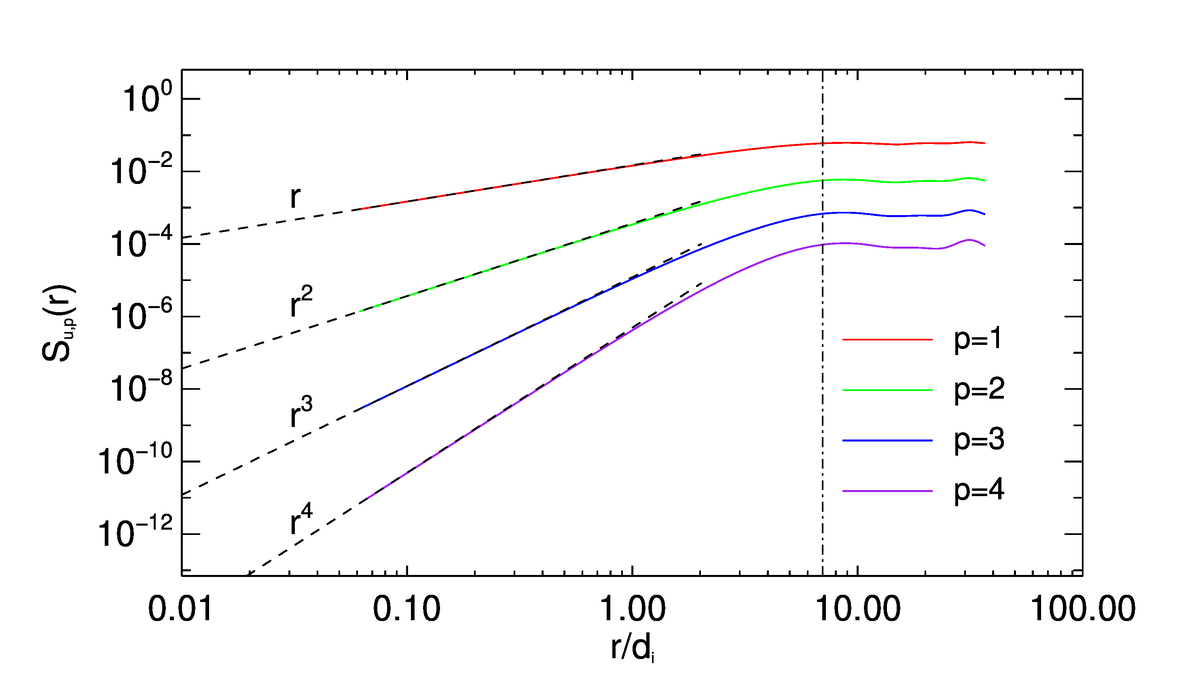} };
\node[] at (-4.1,1.7) { \Large{(a)} };
\end{tikzpicture}
}
\subfloat{
\begin{tikzpicture}
\node[] at (0,0) { \includegraphics[width=.48\linewidth]{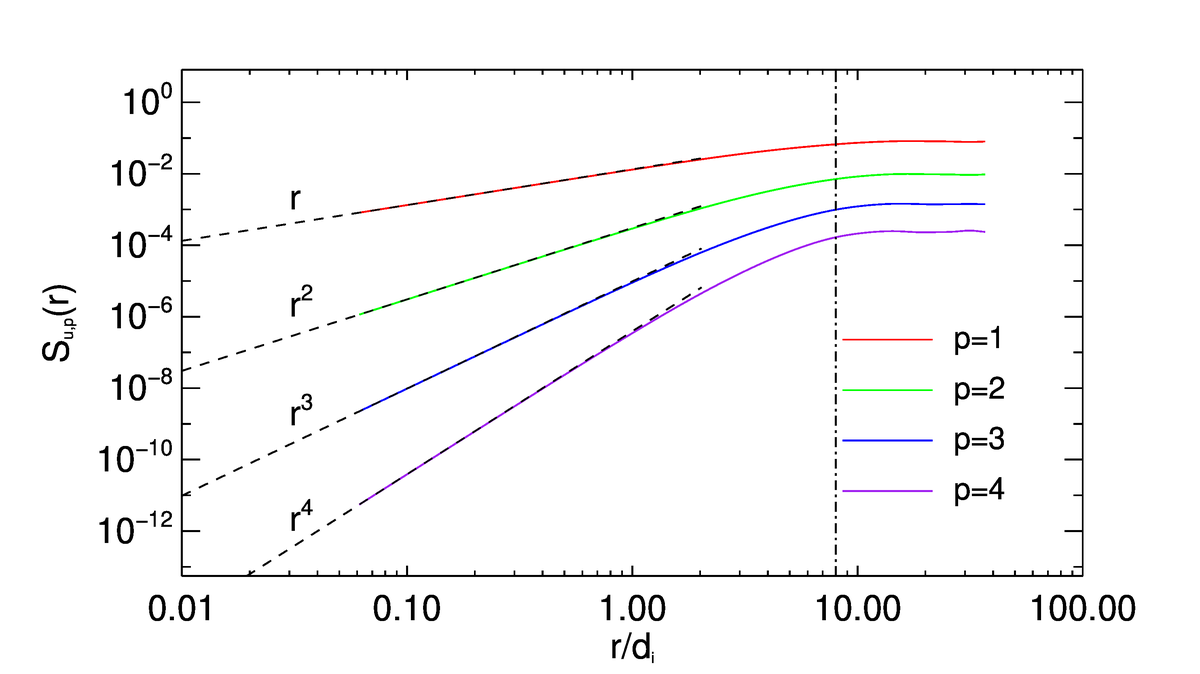} };
\node[] at (-4.1,1.7) { \Large{(b)} };
\end{tikzpicture}
}
\\
\subfloat{
\begin{tikzpicture}
\node[] at (0,0) { \includegraphics[width=.48\linewidth]{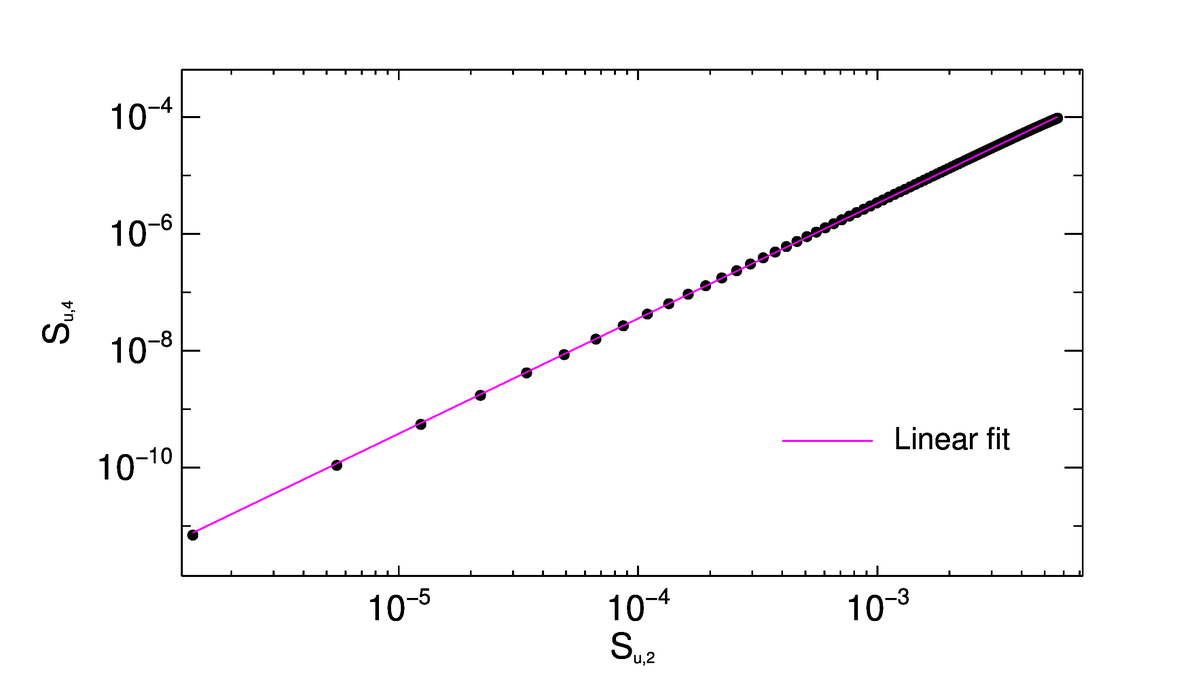} };
\node[] at (-4.1,1.7) { \Large{(c)} };
\end{tikzpicture}
}
\subfloat{
\begin{tikzpicture}
\node[] at (0,0) { \includegraphics[width=.48\linewidth]{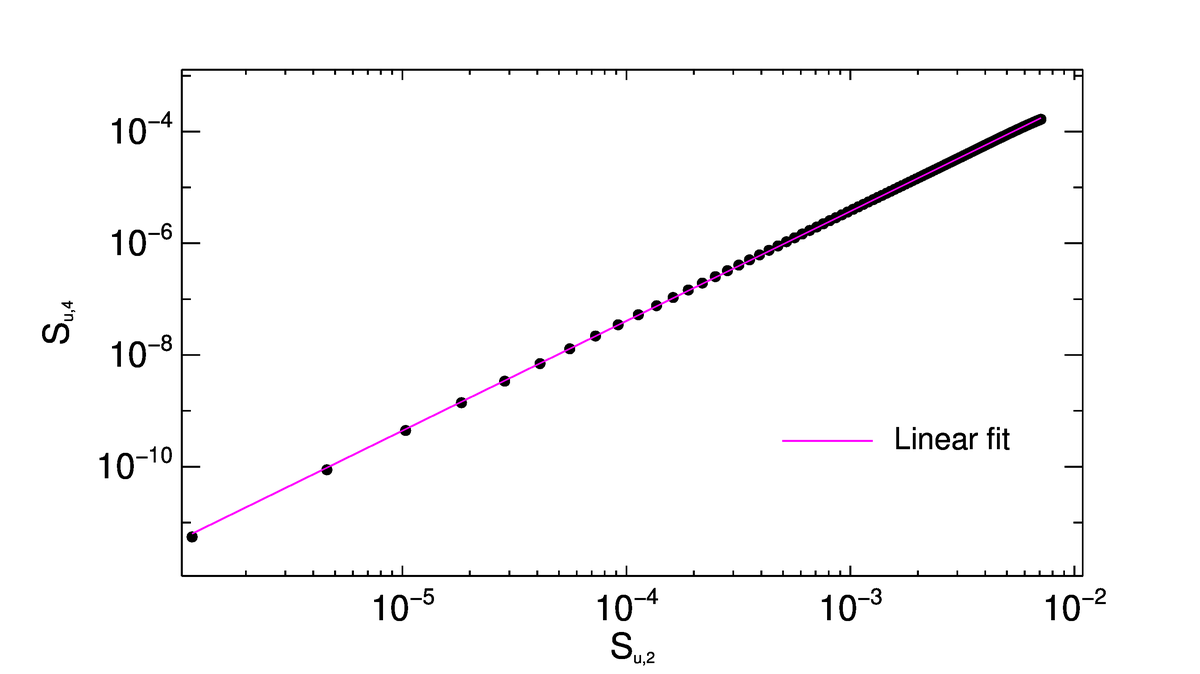} };
\node[] at (-4.1,1.7) { \Large{(d)} };
\end{tikzpicture}
}
\\
\subfloat{
\begin{tikzpicture}
\node[] at (0,0) { \includegraphics[width=.48\linewidth]{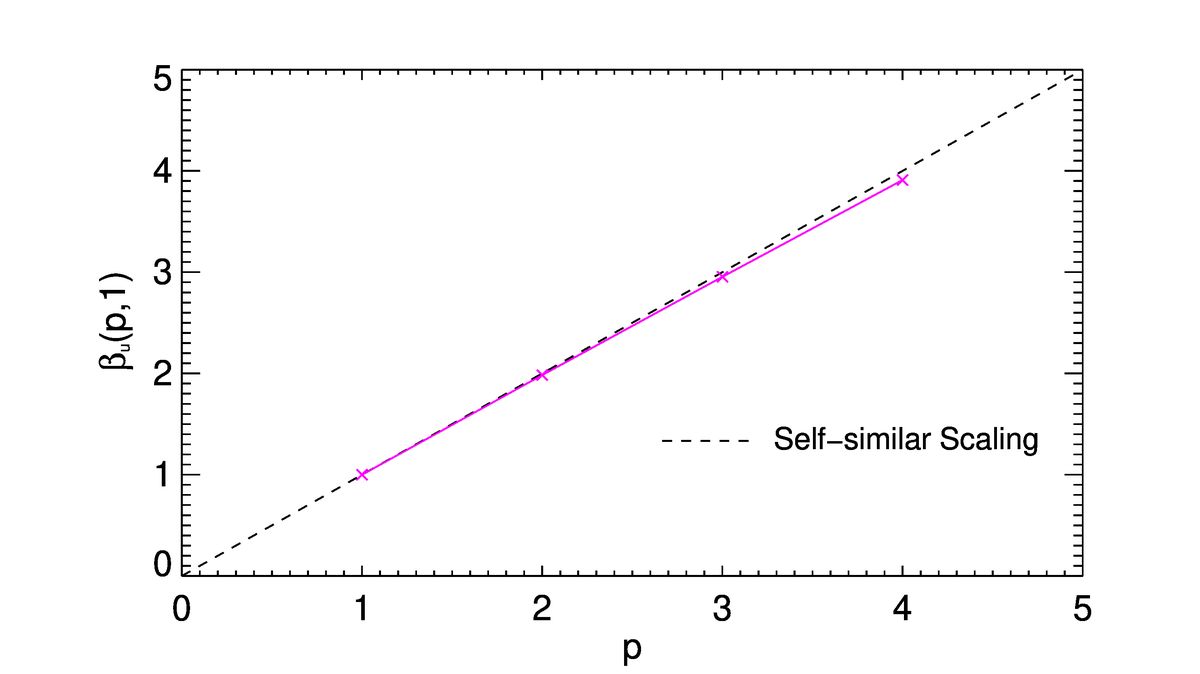} };
\node[] at (-4.1,1.7) { \Large{(e)} };
\end{tikzpicture}
}
\subfloat{
\begin{tikzpicture}
\node[] at (0,0) { \includegraphics[width=.48\linewidth]{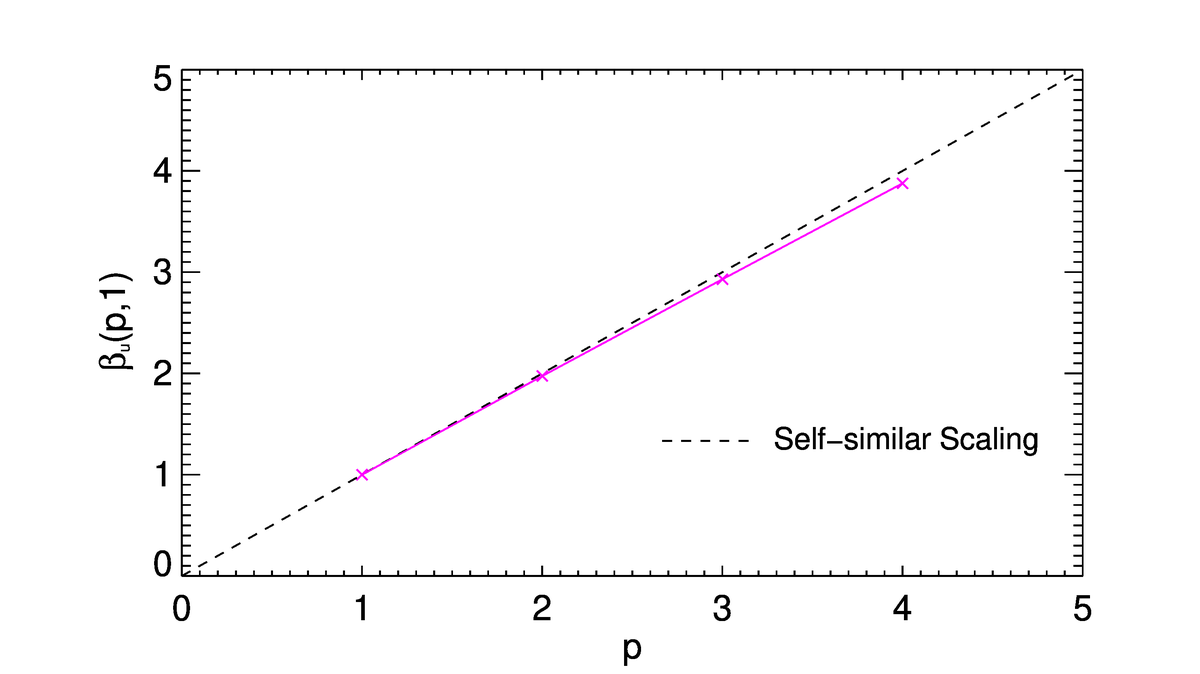} };
\node[] at (-4.1,1.7) { \Large{(f)} };
\end{tikzpicture}
}
\caption{Top panels: Ion velocity structure functions $S_{u,p}$ (in log-scale) of sim.1 at $t_1\!=\!131.7 \,\Omega^{-1}_i$ (a) and of sim.2 at $t_2\!=\!147.5 \,\Omega^{-1}_i$ (b); dashed straight lines represent the power laws $r^p$ , and the vertical dash-dotted lines separate the power law-like region from the saturation region. Middle panels: $S_{u,4}$ vs. $S_{u,2}$ (filled black dots, in log-scale) for $r\!<\!7\,d_i$ from sim.1 at $t_1\!=\!131.7 \,\Omega^{-1}_i$ (c) and sim.2 at $t_2\!=\!147.5 \,\Omega^{-1}_i$ (d); these curves were fit with a straight line (in magenta). Bottom panels: Ion velocity scaling exponents $\beta_u(p,1)$ in range $r\!<\!7\,d_i$ from sim.1 at $t_1\!=\!131.7 \,\Omega^{-1}_i$ (e) and sim.2 at $t_2\!=\!147.5 \,\Omega^{-1}_i$ (f); the dashed straight line, representing the self-similar scaling $\beta(p,1)\!=\!p$, is given as reference.}
\label{fig03}
\end{figure}

The large-scale behavior observed for $r\!>\!10\,d_i$ is expected because we used periodic boundary conditions in a finite box, which causes the SFs to become periodic and even in $r$ \citep{21}. Because of these properties, all SFs here considered tend to grow for $r\!>\!0$ and start to saturate around $r\!=\!L/2$ (where $L$ is the box size), while for $r\!>\!L/2,$ they decrease symmetrically with respect to $r\!=\!L/2$. We did not analyze the SFs for $r\!>\!10\,d_i$ because the statistics there would be affected by these finite-box effects.

The small-scale power law behavior observed in range I is expected as well \citep{18,22} because of the dissipation effects that become important at small $r$ and tend to smooth out magnetic field fluctuations. This implies that in the dissipation range $B_x(x+r,y)-B_x(x,y)\!=\!\Delta B_x(r)\!\sim\!r$ and consequently, the magnetic field SFs take the form of the power law $S_{B,p}\!\sim\!r^p$. This appears evident in the first two panels of figure \ref{fig02}, where in range I all SFs overlap almost perfectly with their corresponding smooth scaling power law $r^p$ in both simulations. Thus, range I can be identified as the magnetic field dissipation range.

As discussed in section \ref{sec04}, even if the SFs do not scale as a power law, as in range II, it can still be possible to characterize the turbulent fluctuations with a set of scaling exponents $\beta(p,q)$ if ESS is observed. Thus, we tested range II for ESS by analyzing all combinations of magnetic field SFs of different order plotted against each other. Panels (c) and (d) of figure \ref{fig02} show an example of two magnetic field SFs of different orders plotted against each other for $r\!<\!10\,d_i$ from sim.1 and sim.2, respectively. Each of these curves was fit separately in range I and range II using two straight lines, and we find that in both simulations, they are linear in range I (blue line) and range II (red line), but with different slopes. This means that ESS holds well in range II but with a different scaling exponent than in range I. The same behavior was found for any other combination of magnetic field SFs of different order in both simulations (not shown here). As ESS is observed, we proceed by evaluating the magnetic field scaling exponents $\beta_B(p,q)$. Panels (e) and (f) of figure \ref{fig02} show $\beta_B(p,q)$ as a function of $p$ at fixed $q\!=\!1$ within range I (blue curve) and range II (red curve) for sim.1 and sim.2, respectively. These exponents were calculated for both ranges separately by taking the gradients of the linear fits of all possible combinations of $log(S_{B,p})$ versus $log(S_{B,q})$. In both simulations, $\beta_B(p,1)$ is linear in $p$ within range I and becomes nonlinear in range II. It is worth noting that a very small deviation from the self-similar scaling is observed in range I as p increases because the calculation of SFs performed by averaging over the simulation grid becomes increasingly less accurate with increasing $p,$ as pointed out in section \ref{sec04}. The same behavior with $\beta_B(p,q)$ being linear in range I and nonlinear in range II was observed for any other value of $q$  in sim.1 and sim.2 (not shown here). This result suggests that in both simulations, the magnetic field fluctuations are intermittent for $r\!>\!0.3\,d_i,$ and they become self-similar for $r\!<\!0.3\,d_i$.

This small-scale transition from an intermittent inertial range to a self-similar magnetic field dissipation range has previously been observed in numerical simulations \citep{12d} and is consistent with the Cluster satellite measurements \citep{12e}. However, in our case, it takes place at scales of about a few electron inertial lengths (around $r\!\simeq\!0.3\,d_i\!\simeq\!3.6\,d_e$) rather than at $r\!\simeq\!1\,d_i$. Furthermore, no relevant differences between the magnetic field statistics of sim.1 and sim.2 are detected, suggesting that the statistical features of the turbulent cascade of magnetic energy, and in particular, the formation of the dissipation range, are independent of the specific reconnection mechanism associated with the evolution of magnetic field fluctuations. These results agree with recent MMS measurements in the magnetosheath of Earth, showing that the statistical properties of turbulent magnetic fluctuations associated with e-rec are analogous to those of other turbulent plasmas where standard reconnection occurs \citep{12c}.

As the magnetic field statistics do not show any significant difference between sim.1 and sim.2, we analyzed the SFs of the ion fluid velocity $\textbf{u}$ to determine whether they show any signature of e-rec because the ions do play a very different role in the reconnection dynamics of the two simulations. In panels (a) and (b) of figure \ref{fig03} we compare the first four ion velocity structure functions $S_{u,p}$ (in logarithmic scale) of sim.1 and sim.2, respectively. All the ion velocity SFs we considered were calculated using $u_x$ , but the same results were obtained using $u_y$ (not shown here). This again implies that ion turbulence is essentially isotropic in our simulations. Surprisingly, as in the case of the magnetic field SFs, figure \ref{fig03} shows that the ion velocity SFs of both simulations shows the same features in the range of scales we considered, and there are no noticeable differences between sim.1 and sim.2. On the other hand, their behavior is significantly different from that of the magnetic field SFs because we do not observe any sub-ion scale transition such as the one between range I and II that characterizes the magnetic field statistics (see the top panels of figure \ref{fig02}). In particular, we see that for $r\!>\!7\, d_i$ all SFs start to saturate, while for $r\!<\!7\, d_i$ , they behave like a power law, although the transition between these two regions is not sharp and introduces some curvature between about $2\, d_i$ and $7\, d_i$.

The large-scale saturation observed for $r\!>\!7\, d_i$ is caused, as in the case of the magnetic field SFs, by the use of periodic boundary conditions in a finite simulation box. We did not analyze the ion velocity SFs in this range because their properties here are significantly affected by these finite box effects.

As concerning the ion velocity SFs behavior for $r\!<\!7\,d_i$, we already said that SFs are usually expected to take the form of the power law $r^p$ at small $r$ because of dissipation that tends to smooth out fluctuations on small scales. However, the first two panels of figure \ref{fig03} show that in both simulations, all ion velocity SFs are well approximated by their corresponding $r^p$ power law for $r\!<\!2\,d_i$, a range that is much wider than the dissipation range of the magnetic field SFs that was identified with range I. This means that the ion velocity fluctuations are smooth on a wider range than the magnetic field fluctuations. However, the formation of this extended ion dissipation range observed in the ion velocity SFs must have a different origin than the magnetic field dissipation range as it covers a range of scales that far exceeds range I and extends to ion scales. A possible explanation is that the development of the ion dissipation range is related to ions being decoupled from the magnetic field at scales of about the ion Larmor radius $\rho_i$ (which is on the same order as $d_i$ for $\beta\!=\!1$, as in our simulations) where ion thermal effects become important. It is reasonable to assume that if the system develops magnetic fluctuations at scales on the same order of $\rho_i$ or smaller, then ions  are unable to follow the rapid magnetic field variations in space, so they will decouple from it and no ion structures will be formed at those scales. Therefore, as an effect of ions decoupling, ion velocity becomes smooth at scales smaller than some $\rho_i\!\simeq\!d_i$. On the other hand, even if ions are decoupled, the intermittent cascade of magnetic energy proceeds toward smaller scales, supported by the electrons that remain coupled to the magnetic field. However, when electron scales are reached, even the electron dynamics decouples from the magnetic field and the magnetic dissipation range is formed. Thus we claim that only the electrons play a role in the formation of the magnetic field dissipation range as the ions decouple from the magnetic field dynamics long before the formation of the magnetic dissipation range.

Furthermore, as all ion velocity SFs exhibit some curvature between $2\,d_i$ and $7\,d_i$, we verified that ESS holds by analyzing all combinations of ion velocity SFs of different order plotted against each other. Panels (c) and (d) of figure \ref{fig03} show an example of two ion velocity SFs of different order plotted against other other for $r\!<\!7\,d_i$ from sim.1 and sim.2, respectively. These curves were fit using a single straight line over the whole range, and we find that in both simulations, they are linear for $r\!<\!7\,d_i$ without any change in slope between the region where all SFs behave like $r^p$ and the region where they show some curvature. This means that ESS holds and that the whole range $r\!<\!7\,d_i$ is characterized by a single scaling exponent. The same behavior was observed for every other combination of ion velocity SFs of different order in both simulations (not shown here). Finally, as ESS is observed, we calculated the ion velocity scaling exponents $\beta_u(p,q)$. Panels (e) and (f) of figure \ref{fig03} show $\beta_u(p,q)$ as a function of $p$ at fixed $q\!=\!1$ for sim.1 and sim.2, respectively. These exponents were calculated taking the gradients of the linear fits of all possible combinations of $log(S_{u,p})$ versus $log(S_{u,q})$. We find that $\beta_u(p,1)$ is linear in $p$ in range $r\!<\!7\,d_i$ , and the same behavior was observed for every other value of $q$  in sim.1 and sim.2 (not shown here). This result suggests that in both simulations, ion velocity fluctuations are self-similar at scales smaller than about $7\, d_i$, even in the region where all SFs show some curvature. The ion velocity fluctuations are therefore likely to be smooth over the whole $r\!<\!7\,d_i$ range.

Thus, the analysis of ion velocity SFs clearly shows that the ion statistics is also not influenced by the specific reconnection mechanism associated with the evolution of magnetic field fluctuations. No signature of e-rec is present because we do not see any difference between the statistical features of the ions in sim.1 and sim.2. 

\section{Conclusions}

By combining the information obtained from the magnetic field and the ion velocity SFs, we find that the turbulent cascade associated with e-rec has the same statistical properties of the turbulent cascade associated with standard reconnection. This result is consistent with a recent analysis of turbulent magnetic fluctuations associated with e-rec, measured in the terrestrial magnetosheath by the satellites of the MMS space mission \citep{12c}.

Furthermore, our analysis suggests that in both simulations, it is possible to identify two dynamical regimes. The first is the ion-decoupled regime, associated with scales in the range $4\,d_e\!<\!r\!<\!7\,d_i$, where magnetic field fluctuations are intermittent while ion velocity fluctuations are self-similar and smooth as this species is strongly decoupled from the magnetic field. The second regime is the dissipative one, associated with scales in the range $r\!<\!4\,d_e$, where both magnetic field and ion velocity fluctuations are self-similar and smooth because of small-scale dissipation. This result is consistent with the analysis of Pyakurel et al. \citep{12b}, according to which ions decouple from the magnetic field at scales of about $10\, d_i\!\simeq\!10\, \rho_i$ and no ion structures are formed at these scales or smaller. In addition, we claim that the formation of the self-similar magnetic field dissipation range is only guided by the small-scale electron dynamics and that it is independent of the ion dynamics as these particles are decoupled from the magnetic field in this range. These results suggest that the statistical features of the turbulent cascade in a collisionless magnetized plasma depend solely on the coupling between the magnetic field and the different particle species present in the system, but they are independent of the specific process that is responsible for the decoupling of these particles (whether it is e-rec or standard magnetic reconnection, e.g.). In other words, this means that e-rec dissipates the turbulent magnetic energy in the same way as standard ion-coupled reconnection does, and this happens because turbulent dissipation is guided by electrons whose dynamics remains unaltered from standard reconnection to e-rec. This seems to be a robust and universal feature of turbulent magnetized plasmas, independent of the reconnection dynamics, and this result has a potential impact on the formulation of new theoretical models of plasma turbulence. In addition, in this context, the SFs proved to be a useful tool for investigating the coupling between particles and the magnetic field, and their use may be extended to the analysis of satellite data as well.

Additional studies are necessary to better characterize the transition between the ion-decoupled regime and the dissipative regime. The spatial grid spacing of our simulations is on the same order as the electron inertial length $d_e$ , and because of this, it is not possible to accurately resolve the small-scale electron dynamics in the dissipation range. Moreover, even if our hybrid model is computationally very efficient and able to highlight the different roles of ions and electrons, it is still too simplified to completely describe the small-scale electron physics. Thus, simulations with higher resolution and including electron kinetic effects are required for a much more detailed study of the formation of the magnetic field dissipation range.

Finally, the natural extension of our work will be to perform full 3D-3V simulations of plasma turbulence to study three dimensional effects on the transition between the different physical regimes that characterize the turbulent energy cascade. 

\section*{\normalsize{Acknowledgments}}

This paper has received funding from the European Unions Horizon 2020 research and innovation programme under grant agreement No 776262 (AIDA, www.aida-space.eu). Numerical simulations have been performed on Marconi at CINECA (Italy) under the ISCRA initiative. FC thank Dr.~M.~Guarrasi (CINECA, Italy) for his essential contribution for his contribution on code implementation on Marconi. We acknowledge dr. S. Cerri, M. Sisti, F. Finelli and S. Fadanelli for very useful discussions.

\section*{\normalsize{Data Availability}}

The simulation dataset (UNIPI e-rec) is available at Cineca on the AIDA-DB. In order to access the meta-information and the link to the raw data, look at the tutorial at http://aida-space.eu/AIDAdb-iRODS.

\bibliography{peppe}

\end{document}